\documentclass[prl,aps,a4paper,showpacs,twocolumn,floatfix]{revtex4}

\usepackage{graphicx}
\usepackage[ansinew]{inputenc}
\usepackage{array}
\usepackage{color}
\usepackage{amsmath}
\usepackage{amsxtra}
\usepackage{amstext}
\usepackage{amssymb}
\usepackage{latexsym}
\usepackage{dsfont}
\begin{document}

\title{Coupling nanomechanical cantilevers to dipolar molecules}

\author{S. Singh, M. Bhattacharya, O. Dutta and P. Meystre}
\affiliation{B2 Institute, Department of Physics and College of Optical
Sciences\\The University of Arizona, Tucson, Arizona 85721}

\date{\today}

\begin{abstract}
We investigate the coupling of a nanomechanical oscillator in the quantum regime with molecular (electric) dipoles. We find theoretically that the cantilever can produce single-mode squeezing of the center-of-mass motion of an isolated trapped molecule and two-mode squeezing of the phonons of an array of molecules. This work opens up the possibility of manipulating dipolar crystals, which have been recently proposed as quantum memory, and more generally, is indicative of the promise of nanoscale cantilevers for the quantum detection and control of atomic and molecular systems.
\end{abstract}


\maketitle
Recent experimental advances have brought macroscopic
oscillators closer than ever before to operating in the
quantum regime \cite{cohadon1999,gigan2006,kleckner2006,
arcizet2006, schliesser2006,corbitt2007}. Technically
progress has been enabled by improvements in nanofabrication and non-equilibrium cooling.
Foundational interest in this frontier lies in the fact
that quantum mechanics has never been tested at such a
macroscopic scale, particularly with respect to
counter-intuitive effects such as superposition and
entanglement. From a practical point of view it is important to explore
the behavior of mechanical oscillators in the quantum
regime since they serve as sensors whose precision is
fundamentally restricted by quantum mechanics
\cite{bocko1996,rugar2004}. 

A broader perspective on the subject may be assumed by taking
into account related successes in atomic physics, where laser
cooling and trapping techniques have enabled impressive
coherent control of microscopic systems. As part of an
ongoing merger between atomic and condensed matter physics,
it has become realistic to explore the interaction of cold
atomic systems with quantum nano-mechanical oscillators. Examples include the coupling of cold ions to vibrating
electrodes \cite{zoller2004}, of a nanomechanical cantilever
to a Bose-Einstein condensate \cite{haensch07}, of an
atomic vapor to an oscillating mirror \cite{genes2008}, etc.

As a first step in this direction, this Letter investigates the
coupling of a laser-cooled nanomechanical cantilever to an
ultracold lattice of polar molecules. Due to the anisotropic,
long-range interaction between these molecules, ensembles of
ultra-cold polar molecules are believed to have a rich phase
diagram and are the subject of intense
theoretical and experimental interest \cite{doyle2004}. The
cantilever-molecule coupling is assumed to be enabled by a
ferroelectric domain mounted on the former leading to a strong
dipole-dipole interaction, and a strong polarizing field freezes
out the rotational freedom of the dipoles.

\begin{figure}[t]
\includegraphics[width=0.35 \textwidth]{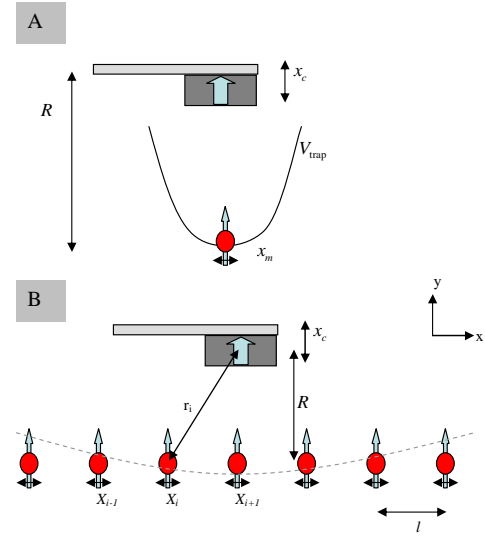}
\caption{\label{fig:setup}(Color online). Arrangement
considered for the coupling a nanomechanical oscillator to a dipolar
``crystal''. In setup A, a single molecule is coupled to the
oscillator. In setup B, the oscillator is
again a distance $R$ from the linear chain of molecules. A weakly confining
harmonic trap for the dipoles is shown along the $x$ axis.}
\end{figure}

To set the stage for the discussion we consider first the simple case of a single molecule, see setup A of Fig.~\ref{fig:setup}, and demonstrate
that its coupling to the cantilever leads to the parametric squeezing of its center-of-mass motion. These considerations are then
generalized to the situation of a linear chain of
electric dipoles (setup B). Such a self-arranged crystal has
recently been proposed as memory for quantum information
processing \cite{rabl2007}. We find that for an appropriate choice of cantilever frequency the phonons in the crystal can be two-mode
squeezed, i.e. entangled \cite{merlin1997,nori1997}, hinting at
the possibility of exploiting such a set-up for the coherent
control of the quantum state of the dipolar lattice. This could for instance be achieved by using a learning algorithm to determine an appropriate time
dependence of the cantilever frequency.

The Hamiltonian describing the coupling of the cantilever to a single molecule is $H = H_c + H_m + V_I $, where
\begin{equation}
\label{eq:NMC}
H_c=\hbar \omega_{c} \left(a^{\dagger}a+\frac{1}{2}\right),
\end{equation}
describes the quantized (single mode of) vibration of the cantilever of
effective mass $m_c$ at frequency $\omega_c$, $a$ and $a^{\dagger}$
being bosonic creation and annihilation operators obeying the
commutation rules $\left[a,a^{\dagger}\right]=1$. In terms of the
displacement $x_c$ of the cantilever along the $y$-axis, we have
\begin{equation}
    x_c=\sqrt{\frac{\hbar}{2 \omega_c m_c}}(a+a^\dagger).
\end{equation}
The Hamiltonian
\begin{equation}
\label{eq:1mol} H_M=\hbar \omega_{t}
\left(b^{\dagger}b+\frac{1}{2}\right),
\end{equation}
describes the center-of-mass motion of the trapped dipole of mass
$m$, where $\omega_t$ is the trap frequency and $b, b^{\dagger}$
are bosonic annihilation and creation operators with
\begin{equation}
    x_m=\sqrt{\frac{\hbar}{2 \omega_t m}}(b+b^\dagger),
\end{equation}
 $x_m$ being the displacement of the molecule along the $x$-axis.
 Finally, the interaction between the molecule and the oscillator
is given by
\begin{equation}
\label{eq:1int}
V_I= \frac{d_{m}d_{c}}{4 \pi\epsilon_0 r^{3}}\left[1-\frac{3 (R+x_c)^2}
{r^{2}}\right],
\end{equation}
where $d_c$ is the dipole moment of the ferroelectric domain
attached to the tip of the nanomechanical cantilever and $R$ is its
distance from the equilibrium position of the molecule. The
distance between the cantilever and the molecule is $r=\left[
(R+x_c)^2+x_m^2\right]^{1/2}$. For $R \gg x_m,x_c$ the dipolar
interaction can be approximated as:
\begin{equation}
\label{eq:Vint1}
 V_I \approx \frac{d_m d_c}{2 \pi \epsilon_0 R^6}(-R^3
 + 3x_cR^2+3x_m^2R- 15x_cx_m^2).
\end{equation}
The typical trap level spacing is much larger than the thermal energy of the ultracold molecule, which justifies its zero-temperature description. We will be including thermal effects for the cantilever below.

The presence of the cantilever has two major effects on the molecule dynamics. First, it leads to a tightening of the trap for small distances $R$, resulting in a shifted trapping frequency
\begin{equation}
 \omega_{t}' = \left [\omega_t^2+\frac{3d_m d_c}{\pi \epsilon_0 m
 R^5}\right ]^{1/2}.
\end{equation}
The second, more interesting effect is parametric squeezing. In an interaction
picture with respect to the free Hamiltonian $H_c+H_m$, taking
$\omega_c = 2 \omega_{t}'$, performing the rotating-wave
approximation, and further assuming that the cantilever motion can
be treated classically,  $(a \rightarrow \alpha)$, the interaction
potential $V_I$ reduces to
\begin{equation}
 \label{eq:oms}
V_I= - \hbar C\left( b^2 + b^{\dagger 2}\right)
\end{equation}
where
\begin{equation}
 C=  \sqrt{\bar{N}}\frac{15d_m d_c}{4\pi\epsilon_{o} m \omega_{t}'R^{6}}\left(\frac{\hbar}{2m_c\omega_c}\right)^{1/2}.
\end{equation}
Here $\bar{N}$ is the average number of quanta of excitation of the mechanical oscillator, $\bar{N}=k_BT_c/\hbar\omega_c$, where $k_B$ is the Boltzmann constant and $T_c$ the cantilever temperature.
Eq.~(\ref{eq:oms}) is the standard squeezing Hamiltonian
familiar from studies of the degenerate parametric amplifier in
quantum optics, see e.g. Ref.~\cite{zubairy1983}.

Thermal noise can be introduced simply in the description of the system in the form of phase fluctuations in the cantilever field. These fluctuations are related to the cantilever damping rate,
$D$ by the fluctuation-dissipation theorem. For times t such that $D < t^{-1}<2 C$, the variance in the dimensionless position quadrature for the molecule, $x_1= \frac{1}{2}(b+b^{\dagger})$, is then given by \cite{zubairy1983}
\begin{equation}
 (\Delta x_1)^2_t = \frac{1}{4}e^{-2u}+\frac{1}{8}e^{2u}Dt
\end{equation}
where $u=2Ct$ is the squeezing parameter.

\begin{figure}[t]
\includegraphics[width=0.42 \textwidth]{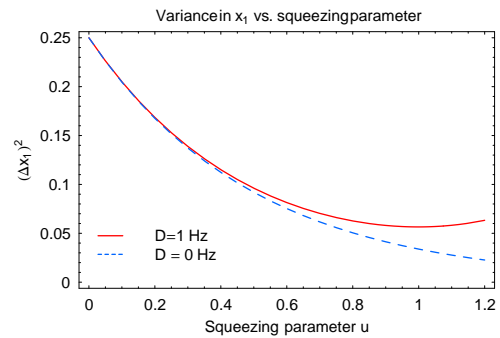}
\caption{\label{fig:squeeze1mode} Variance in the quadrature component $x_1$ (Eq. 12), vs. squeezing parameter $u$, for a SrO molecule interacting with a cantilever. The two curves are for the cantilever damping rates linewidth D = 1 Hz (solid curve) and 0 Hz (dashed curve). Squeezing occurs when the variance in $x_1$ is below $1/4$, and is eventually destroyed due to phase noise in the cantilever.}
\end{figure}

Consider for example a nanomechanical cantilever with frequency
$\omega_c = 4$MHz, effective mass $m_c = 10^{-16}$kg, and
linewidth $D = 1$Hz. A ferroelectric domain with dipole moment
$d_c = 2.1 \times 10^{-23}$C-m is attached to the cantilever and is
placed at $R = 2\mu$m from a SrO molecule. These parameters give
an oscillator frequency $\omega_t' = 2$MHz. We assume $\bar{N}=100$, yielding $C = 20.4$ Hz. Fig.~\ref{fig:squeeze1mode} shows the
variance in $x_1$ as a function of the squeezing parameter for
that example. We remark that the squeezing in single trapped ions \cite{meekhof96} and atoms \cite{raithel97} has been experimentally demonstrated, and similar measurement techniques techniques can possibly be implemented to detect squeezing in the present case.

We now extend these considerations to the case of a lattice of $N$ heteronuclear molecules contained in a harmonic trap $V_{t}$, as shown in Fig. 1.B. The trap is arranged so as to confine the sample weakly along $x$,
tightly along $y$ and tightly or weakly along the $z$
direction depending on whether a one- or two-dimensional lattice
crystal is desired. (We restrict our considerations to the case of a one-dimensional chain in the following.) A polarizing electric field is
provided along $y$ so that all the dipoles align along that direction. The system can be described by the Hamiltonian
\begin{equation}
 \label{eq:classphon}
H_p = \displaystyle \sum_{i}^{N} \frac{p_{i}^{2}}{2m}
+\frac{d_{m}^{2}}{4\pi\epsilon_{o}} \displaystyle \sum_{i < j}^{N}
\frac{1}{\left|x_{i}-x_{j}\right|^3}+V_{t},
\end{equation}
where $x_{i},p_{i}$ are the position and momentum, of the $i$th molecule
and $V_{t}$ is the external trapping potential. The first term in
Eq.~(\ref{eq:classphon}) corresponds to the kinetic
energy of the dipoles, the second to their dipole-dipole
interaction and the last term denotes the trap energy.

Due to their mutual repulsion along the $x$ direction
the molecules self-organize into a linear lattice \cite{rabl2007}.
For small oscillations of the molecules about their
equilibrium positions Eq.~(\ref{eq:classphon}) can be
expressed in terms of acoustic phonon modes of momentum
$k$ and energy $\hbar \omega_{k}$,
\begin{equation}
 \label{eq:phonons}
H_p = \displaystyle \sum_{k} \hbar \omega_{k} \left(b^{\dagger}_{k}b_{k}
+\frac{1}{2}\right),
\end{equation}
where $b_{k}, b^{\dagger}_{k}$ are the phonon creation and
annihilation operators obeying the bosonic commutation rules
$\left[b_{k},b_{k'}^{\dagger}\right]=\delta_{kk'}$,
the phonon frequencies are given by $\omega_{k}=2\omega_{o}\left|\sin(kl/2)\right|$, where $\omega_{o}=d_{m}\left(3/2\pi\epsilon_{o}ml^{5}\right)^{1/2}$,
and $l$ is the lattice spacing. We note that only terms harmonic in the $x_i$ have been retained in deriving Eq.~(\ref{eq:phonons}) from Eq.~(\ref{eq:classphon}). Higher, anharmonic terms represent phonon-phonon interactions and in particular determine the phonon lifetime in the crystal \cite{pathak65}.

We consider ultracold molecules at a temperature $T$ such that $k_{B}T \ll \hbar \omega_{o},$ so that a $T=0$ description is appropriate as before for the molecules. The energy of the nanomechanical cantilever is again given by Eq.~(\ref{eq:NMC}), so that the interaction between the chain of molecules and the oscillator is \begin{equation}
\label{eq:int}
V_I= \displaystyle \sum_{i}\frac{d_{m}d_{c}}{r_{i}^{3}}\left[1-\frac{3 (R+x_c)^2}
{r_{i}^{2}}\right],
\end{equation}
where $d_c$ is the dipole moment of the ferroelectric domain
attached to the tip of the cantilever. Here $x_c$ is the
displacement of the cantilever along the $y$ axis, $R$ is
its distance from the center of the dipolar crystal and its
distance from the $i$th molecule is given by $r_{i}=\left[
(R+x_c)^2+(il+x_{i})^2\right]^{1/2}$. Exploiting
the hierarchy of length-scales, $x_{i} \ll l, Nl \ll R$,
we expand Eq.~(\ref{eq:int}) to find that the oscillator produces
a slight shift in the phonon frequency,
\begin{equation}
\omega_{k}'\approx \omega_{k}+\frac{d_{m}d_{c}}{4\pi\epsilon_{o}m \omega_{k}^{2}R^{5}}.
\end{equation}
and the coupling of the cantilever to the phonons is given by
\begin{equation}
V_{I}=-\displaystyle \sum_{k} \hbar C_{k}' \left(a+a^{\dagger}\right)
\left(b_{k}b_{-k}+b_{k}^{\dagger}b_{-k}^{\dagger}+b_{k}^{\dagger}
b_{k}+b_{-k}b_{-k}^{\dagger}\right),
\end{equation}
where
\begin{equation}
 \label{eq:ckp}
C_{k}'=-\frac{3d_m d_c}{2\pi\epsilon_{o} m \omega_{k}'R^{6}}\left(\frac{\hbar}{2m_c\omega_c}\right)^{1/2}.
\end{equation}
We remark that for sufficiently small $R$ and/or sufficiently large $d_c$
the cantilever will couple to the individual dipoles rather than collectively to the acoustic phonons.

Following an approach that parallels the single-molecule description we work in an interaction picture with respect to the free Hamiltonian of the cantilever and the (frequency shifted) phonon mode. We further choose the cantilever frequency such that $\omega_c=2\omega_{k}'$, with $k=\pi/l$, implying that the cantilever couples primarily to excitations near the edge of the first Brillouin zone, where the density of phonon states is largest. We assume for simplicity that the motion of the nanomechanical cantilever can be described classically, $(a \rightarrow \alpha)$, a reasonable approximation since it is still challenging to cool these systems to their quantum regime. Performing the rotating wave approximation we then obtain the approximate interaction picture interaction Hamiltonian
\begin{equation}
 \label{eq:tms}
V_I= - \hbar C_{k}\left( b_{k}b_{-k}
+ b_{-k}^{\dagger}b_{k}^{\dagger}\right).
\end{equation}
where $C_k =\sqrt{\bar{N}} C_k'$ and $\bar{N}$ is the average occupation number of the cantilever. This Hamiltonian is known from quantum optics to lead to the generation of two-mode squeezing between acoustic phonon modes of momenta  $\pm k$ (within the bandwidth of the nanomechanical resonance), and hence their quantum entanglement. 

\begin{figure}[t]
\includegraphics[width=0.42 \textwidth]{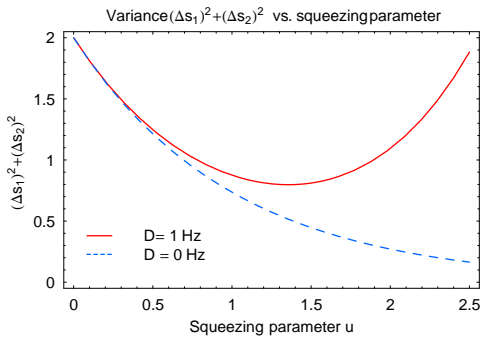}
\caption{\label{fig:squeeze2mode} Sum of variances of $s_1$ and $s_2$, vs. squeezing parameter $u$, for a SrO dipolar crystal interacting with a cantilever. The two curves are for cantilever damping rates of D = 1 Hz (solid curve) and 0 Hz (dashed curve). Phase fluctuations of the cantilever eventually increase the sum of variances to a value larger than 2, indicative of the loss of entanglement between the two phonon modes.}
\end{figure}

In order to characterize that two-mode squeezing we introduce the two dimensionless quadratures as follows:
\begin{eqnarray}
 s_1= \frac{1}{\sqrt{2}}(b_k+b_{-k}+b_k^{\dagger}+b_{-k}^{\dagger})\\
 s_2=\frac{1}{\sqrt{2}i}(b_k-b_k^{\dagger}-b_{-k}+b_{-k}^{\dagger})
\end{eqnarray}
Taking into account the phase fluctuations in the cantilever motion resulting from thermal noise, the sum of variances in the two quadratures is then \cite{zubairy2006}:
\begin{eqnarray}
&&\lefteqn{(\Delta s_1)^2+(\Delta s_2)^2= } \nonumber\\
 & & \frac{e^{-\frac{Dt}{2}}}{C_{k0}}\{ D\sinh(C_{k0}t)+ 2C_{k0} \cosh (C_{k0}t)\} \nonumber \\
 & & -\sum_{i,j,k,i\neq j,j\neq k} e^{\lambda_it}\frac{2 C_{k0}(\lambda_i+4D)}{(\lambda_i-\lambda_j)(\lambda_i-\lambda_k)},
\end{eqnarray}
 where $u$ is the squeezing parameter and is equal to $2C_{k0}t$, with $C_{k0} = \frac{1}{2}\sqrt{4C_k^2-D^2}$ and the $\lambda_i$'s are the roots of the cubic equation:
\begin{equation}
  \lambda^{3} + 5D\lambda^{2}+(4D^2-C_{k0}^2)\lambda - 2C_{k0}^2 D = 0
\end{equation}
For example, let us consider a nanomechanical cantilever with frequency $\omega_c = 2$MHz, effective mass $m_c = 10^{-16}$kg, and linewidth $D = 1$Hz. We assume an average occupation number of 100. A ferroelectric domain with dipole moment $d_c = 2.1 \times 10^{-23}$C-m is attached to the cantilever and is placed at $R = 2\mu$m from a one dimensional SrO crystal. The crystalline phase of these dipolar molecules is formed with inter-molecular distances $l \approx 200$nm. These parameters give a phonon frequency $\omega_o = 4$MHz, and thus an interaction $C_k = 6.2$. The squeezing parameter $u$ is given by $2C_kt$. Fig. \ref{fig:squeeze2mode} gives the sum of variances of $s_1$ and $s_2$ as a function of the squeezing parameter for this system.

The sum of the variances in Eq. 20 is actually a measure of the entanglement of the system \cite{zoller2000}. In our case, this inseparability criterion implies that the system is entangled if the sum of the variances is less than 2. We observe from Fig. 3 that as expected intuitively the introduction of phase fluctuations destroys the entanglement over time. As far as the experimental verification of this prediction is concerned we remark that squeezed states of phonons have been previously detected experimentally in solid-state systems \cite{merlin1997}, and similar techniques can possibly be used to detect squeezed phonon modes in the present system.

As indicated earlier, our results assume that the motion of the cantilever is classical and dominated by thermal rather than quantum fluctuations. In addition to being a realistic description of the current experimental situation, this approximation enabled us to present the physics of the coupling between the cantilever and the dipolar molecules using simple analytical models. However, our results are expected to still hold at least qualitatively when treating the cantilever quantum-mechanically. Exact numerical solutions for one-mode and two-mode squeezing using a coherent pump with low $\bar{N}$ are available in literature \cite{kinsler93,crouch88}. These results are consistent with our simple analytical treatment and point to the existence of squeezing for this system.

In conclusion, we studied the coupling of nanomechanical cantilevers to dipolar molecules. We found that for a single trapped molecule, the presence of the cantilever leads to tighter confinement and parametric squeezing. We also demonstrated squeezing and entanglement of the phonon modes of a linear chain of dipolar molecules. These results open up the way to extremely promising novel methods for the quantum manipulation and control of the state of ultracold dipolar systems, and are indicative of the general use of nanoscale cantilevers in the detection and control of atomic and molecular systems.

This work is supported in part by the US Office of
Naval Research, by the National Science Foundation,
and by the US Army Research Office.



\begin{thebibliography}{10}

\bibitem{cohadon1999}
P. F. Cohadon, A. Heidmann and M. Pinard, Phys. Rev. Lett. {\bf 83},  3174  (1999).

\bibitem{gigan2006}
S. Gigan \textit{et al.}, Nature {\bf 444},  67  (2006).

\bibitem{kleckner2006}
D. Kleckner and D. Bouwmeester, Nature {\bf 444},  75  (2006).

\bibitem{arcizet2006}
O. Arcizet, P. -F. Cohadon, T. Briant, M. Pinard, and A. Heidmann, Nature {\bf 444},  71
  (2006).

\bibitem{schliesser2006}
A. Schliesser \textit{et al.}, Phys. Rev. Lett. {\bf 97},  243905  (2006).

\bibitem{corbitt2007}
T. Corbitt \textit{et al.}, Phys. Rev. Lett. {\bf 98},  150802  (2007).

\bibitem{bocko1996}
M. Bocko and R. Onofrio, Rev. Mod. Phys. {\bf 68},  755  (1996).

\bibitem{rugar2004}
D. Rugar, R. Budakian, H. J. Mamin and B. W. Chui, Nature {\bf 430},  329  (2004).

\bibitem{zoller2004}
L. Tian and P. Zoller, Phys. Rev. Lett. {\bf 93},  266403  (2004).

\bibitem{haensch07}
P. Treutlein  \textit{et. al}, Phys. Rev. Lett. {\bf 99},  140403  (2007).

\bibitem{genes2008}
C. Genes, D. Vitali, P. Tombesi, arXiv:0801.2266  (2008).

\bibitem{doyle2004}
J. M. Doyle, B. Friedrich, R. V. Krems, and F. Masnou-Seeuws, Eur. Phys. J. D {\bf
  31},  149  (2004).

\bibitem{rabl2007}
P. Rabl and P. Zoller, Phys. Rev. A {\bf 76},  042308  (2007).

\bibitem{merlin1997}
G. A . Garrett, A. G. Rojo, A. K. Sood, J. F. Whitaker and R. Merlin, Science {\bf 275},
  1638  (1997).

\bibitem{nori1997}
X. Hu and F. Nori, Phys. Rev. Lett. {\bf 79},  4605  (1997).

\bibitem{zubairy1983}
K. Wodkiewicz and M.~S. Zubairy, Phys. Rev. A {\bf 27},  2003  (1983).

\bibitem{meekhof96}
D.M. Meekhof, C. Monroe, B.E. King, W.M. Itano and D.J. Wineland, Phys. Rev. Lett. {\bf 76},
   1796  (1996).

\bibitem{raithel97}
G. Raithel, G. Birkl, W.D. Phillips, and S.L. Rolston, Phys. Rev. Lett. {\bf 78},  2928
  (1997).

\bibitem{pathak65}
K. Pathak, Phys. Rev. {\bf 139},  1569  (1965).

\bibitem{zubairy2006}
K. Ahmed, H. Xiong and M.S. Zubairy, Opt. Comm. {\bf 262},  129  (2006).

\bibitem{zoller2000}
L-M. Duan, G. Giedke, J.I. Cirac and P. Zoller, Phys. Rev. Lett. {\bf 84},  2722
  (2000).

\bibitem{kinsler93}
P. Kinsler, M. Fernee and P.D. Drummond, Phys. Rev. A {\bf 48},  3310  (1993).

\bibitem{crouch88}
D. Crouch and S. Braunstein, Phys. Rev. A {\bf 38},  4696  (1988).

\end{thebibliography}
\end{document}